%% file: main.tex
\documentclass[11pt]{article}

\usepackage[margin=1in]{geometry}
\usepackage{amsmath,amssymb,amsthm}
\usepackage{graphicx}
\usepackage{bm}
\usepackage{hyperref}
\usepackage{xcolor}
\input{macros}

\title{
Comparing Implicit Neural Representations and B-Splines\\
for Continuous Function Fitting from Sparse Samples
}

\author{
Hongze Yu$^{1}$,
Yun Jiang$^{2,3}$,
Jeffrey A.\ Fessler$^{1,2,3}$,
\\[0.4em]
{\small $^{1}$Department of Electrical Engineering and Computer Science, University of Michigan}\\[-0.2em]
{\small $^{2}$Department of Biomedical Engineering, University of Michigan}\\[-0.2em]
{\small $^{3}$Department of Radiology, University of Michigan}
}

\date{\today}

\begin{document}
\maketitle

\begin{abstract}
Continuous signal representations are naturally suited for inverse problems,
such as magnetic resonance imaging (MRI) and computed tomography, 
because the measurements depend on an underlying physically continuous signal. 
While classical methods rely on predefined analytical bases like B-splines, 
implicit neural representations (INRs) have emerged as a powerful alternative 
that use coordinate-based networks to parameterize continuous functions with 
implicitly defined bases. Despite their empirical success, direct comparisons 
of their intrinsic representation capabilities with conventional models 
remain limited. This preliminary empirical study compares a positional-encoded INR 
with a cubic B-spline model for continuous function fitting from sparse random samples, 
isolating the representation capacity difference by only using coefficient-domain 
Tikhonov regularization. Results demonstrate that, under oracle hyperparameter selection, 
the INR achieves a lower normalized root-mean-squared error, yielding sharper edge 
transitions and fewer oscillatory artifacts than the oracle-tuned B-spline model. 
Additionally, we show that a practical bilevel optimization framework for INR hyperparameter 
selection based on measurement data split effectively approximates oracle performance. 
These findings empirically support the superior representation capacity of INRs for sparse data fitting.
\end{abstract}

\section{Introduction}


Continuous signal representations model an unknown function $f(\vecb r)$ 
using a parametric mapping $f_{\bc}(\vecb r)$, where the coefficients $\bc$ are estimated 
from sampled data.
The resulting function can be evaluated at arbitrary spatial locations.
This approach is naturally suited for inverse problems,
such as magnetic resonance imaging (MRI)~\cite{MBIR_MRI}
and computed tomography,
because the recorded measurements
(k-space samples in MRI and sinograms in tomography)
depend on an underlying signal that is physically continuous. 
Classical formulations typically use approximation spaces
defined by analytical bases or kernels,
including Fourier expansions,
splines, and radial basis functions%
~\cite{wahba1990spline, splineSPIM, mallat1999wavelet, wendland2005scattered}. 
Among these, B-splines are a common choice, where $f(\vecb r)$ is represented
in a finite-dimensional spline space defined by a set of knots.
The knot density balances between the approximation error 
and the noise sensitivity.
In practical applications,
the sensing matrix derived from basis functions
is often ill-conditioned due to sparse sampling.
A standard approach is to regularize the inverse problem with
handcrafted smoothness- or sparsity-promoting penalties.

Recent advances in computer graphics and computer vision,
especially in neural rendering and view synthesis~\cite{nerf},
have boosted the development of various continuous signal representations. 
Explicit models construct scenes leveraging localized data structures or 
graphic primitives, such as interpolated feature 
grids~\cite{fridovich2022plenoxels} and 3D splats based on probability distributions~\cite{kerbl3Dgaussians}. 
In contrast, implicit models represents the signal as a continuous, 
differentiable function that can be queried at arbitrary coordinates%
~\cite{park2019deepsdf,nerf, barron2021mipnerf,instantngp, FourierFeautreNet}.
Among these, implicit neural representations (INRs)~\cite{siren}
have emerged as a powerful class of continuous parametric models 
in which a multilayer perceptron (MLP), composed with a positional encoder,
maps spatial coordinates $\vecb{r}$ directly to signal values.
Unlike the predefined analytical bases in conventional methods,
the effective basis of an INR is implicitly defined
by the neural network architecture and its weights. 
INRs have become increasingly popular in MRI%
~\cite{NeRP, IMJENSE, spatiotemporal_inr, bilevelINR_arXiv, lyu2026rapid} in tasks such as undersampled reconstruction and super-resolution,
where the continuous formulation aligns with the underlying physics of MRI. 

Despite their empirical success,
few works have directly compared conventional analytical methods
(e.g., B-splines) and INRs in terms of their representation capabilities.
To address this gap and motivate the use of INRs, 
this paper compares INRs with B-spline models
for continuous function fitting from sparse, randomly sampled data.
In this preliminary empirical study,
both methods employed Tikhonov regularization on their weights or coefficients,
providing a direct comparison of their intrinsic representation capacities,
without confounding influence of additional handcrafted regularizers.
Furthermore, we investigate hyperparameter selection in this function-fitting task. For B-splines,
we perform an oracle search (assuming access to ground truth)
to determine the best possible performance. For INRs, we use a practical 
bilevel optimization approach~\cite{bilevelINR_arXiv}, which relies on a train/validation split only..

The remaining materials are organized as follows.
Section~\RNum{2} details the problem formulation and the compared methods.
Section~\RNum{3} describes the experiment settings.
Sections~\RNum{4} and~\RNum{5} present experiment results and discussions.

\section{Methods}

\subsection{Problem Formulation}

Let $f(\vecb{r}): \reals^d \rightarrow \reals$
denote a continuous target function defined on a spatial domain.
We observe $N$ samples at random locations 
$\{\vecb{r}_n\}_{n=1}^{N}$
with corresponding values
$\{y_n\}_{n=1}^{N}$, 
where $y_n = f(\vecb{r}_n)$.
The goal is to recover $f$ on an arbitrary dense evaluation grid
$\ztilt \in \reals^{N_{\mathrm{eval}} \times d}$
from the measured vector
$\y = (y_1,\ldots,y_N)^T$.

\subsection{B-Spline Formulation}

As a method that is representative
of typical predefined spatial basis functions
using in solving inverse problems,
we investigated cubic B-splines with $M$ uniformly spaced knots per dimension,
yielding $M+2$ basis functions per dimension and $(M+2)^2$ coefficients 
for the 2D function approximation using the standard tensor product representation:
\begin{equation}
  f_{\bc}(\vecb{r}) 
  = \sum_{i,j=1}^{M+2} 
    c_{ij} \,
    \beta^{(3)}\!\paren{\frac{r_x - \xi_i}{\Delta}} \,
    \beta^{(3)}\!\paren{\frac{r_y - \eta_j}{\Delta}},
  \label{eq:bspline_2d}
\end{equation}
where $\Delta$ is the knot spacing,
and $\bc \in \reals^{(M+2)^2}$ denotes 
the vectorized coefficients to be computed from \y.

Hence, the Tikhonov regularized B-spline fitting is formulated as:
\begin{equation}
\hat{\bc}(\lambda) = \argmin_{\bc} \|\y - \bPhi \bc\|_2^2 + \lambda \|\bc\|_2^2,
\label{eq:bspline_opt}
\end{equation}
where $\bPhi$ is the design matrix (sensing matrix)
of B-spline basis functions
evaluated at the sample locations.
This quadratic cost function has a closed-form minimizer
that we compute using Cholesky decomposition
\begin{equation}
\hat{\bc}(\lambda)
= \paren{\bPhi^\top \bPhi + \lambda \bI}^{-1}\bPhi^\top \y.
\label{eq:bspline_closed_form}
\end{equation}

\subsection{INR Formulation}

The INR parameterizes $f$ as a composition of a positional encoder 
and an MLP decoder:
\begin{equation}
f_{\btheta}(\vecb{r}) = M_{\btheta_{\mathrm{MLP}}}\paren{\gamma_{\btheta_{\mathrm{Enc}}}(\vecb{r})},
\label{eq:inr_representation}
\end{equation}
where $\gamma_{\btheta_{\mathrm{Enc}}}: \reals^d \rightarrow \reals^{KF}$ is 
the multiresolution hash encoder with $K$ levels and $F$ features per level, 
and $M_{\btheta_{\mathrm{MLP}}}: \reals^{KF} \rightarrow \reals$ is the MLP decoder.
The network weights $\btheta$ include both encoder weights $\btheta_{\mathrm{Enc}}$ 
and decoder weights $\btheta_{\mathrm{MLP}}$.

The weight-regularized INR fitting is formulated as:
\begin{equation}
\hat{\btheta}(\bbeta) = \argmin_{\btheta}
\frac{1}{N}
\|\y - f_{\btheta}(\z)\|_2^2 + \lambda_{\mathrm{Enc}} \|\btheta_{\mathrm{Enc}}\|_2^2 + \lambda_{\mathrm{MLP}} \|\btheta_{\mathrm{MLP}}\|_2^2,
\label{eq:inr_opt}
\end{equation}
where $\z \in \reals^{N \times d}$ denotes the sample locations and $\y \in \reals^N$ 
the corresponding observations. The hyperparameter vector 
$\bbeta = (\lambda_{\mathrm{Enc}}, \lambda_{\mathrm{MLP}}, \tau, b)$ includes the 
regularization strengths for the encoder and MLP weights, learning rate $\tau$
and the cross-level resolution scale factor $b$ for the encoder.
We use Adam~\cite{adam} to perform the minimization in \eqref{eq:inr_opt}.

\subsection{Oracle Hyperparameter Search for B-Spline}

For B-spline fitting, we perform an oracle grid search
over the regularization strength $\lambda$ and number of knots $M$,
selecting the hyperparameters that minimize the
normalized root-mean-squared error (NRMSE) on the dense evaluation grid $\ztilt$:
\begin{equation}
\lambda^*, M^* = \argmin_{\lambda, M}\
\frac{\left\| f_{\hat{\bc}(\lambda, M)}(\ztilt) - f_{\mathrm{true}}(\ztilt) \right\|_2}
{\left\|f_{\mathrm{true}}(\ztilt)\right\|_2}.
\label{eq:bspline_oracle}
\end{equation}
This oracle search assumes access to the ground truth
and serves to bound the achievable performance.

\subsection{Bilevel Optimization for INR}

In practice, the ground truth is unavailable, and performing an oracle search for a
multidimensional hyperparameter vector can be computationally inefficient.
Following the self-supervised hyperparameter optimization framework in~\cite{bilevelINR_arXiv},
we split the observed samples into a training set \cT and a validation set \cV,
and performed bilevel optimization for INR:
\begin{equation}
\begin{aligned}
\bbeta^* &= \argmin_{\bbeta} \mathcal{L}_{\cV}\paren{\hat{\btheta}(\bbeta)}, \\
\text{s.t.} \quad \hat{\btheta}(\bbeta) &= \argmin_{\btheta} \mathcal{L}_{\cT}(\bbeta, \btheta),
\end{aligned}
\label{eq:inr_bilevel}
\end{equation}
where
\begin{align}
\mathcal{L}_{\cV}\paren{\hat{\btheta}(\bbeta)}
&= \frac{1}{|\cV|} 
\sum_{n \in \cV} \paren{y_n - f_{\hat{\btheta}(\bbeta)}(\vecb{r}_n)}^2,
\label{eq:inr_val_loss} \\
\mathcal{L}_{\cT}(\bbeta, \btheta)
&= \frac{1}{| \cT |}
\sum_{n \in \cT} \paren{y_n - f_{\btheta (\bbeta)}(\vecb{r}_n)}^2
+ \lambda_{\mathrm{Enc}} \|\btheta_{\mathrm{Enc}}\|_2^2
+ \lambda_{\mathrm{MLP}} \|\btheta_{\mathrm{MLP}}\|_2^2, \label{eq:inr_train_loss}
\end{align}
and $\bbeta = (\lambda_{\mathrm{Enc}}, \lambda_{\mathrm{MLP}}, \tau, b)$
includes regularization strengths, learning rate, and encoder scale factor.

After computing $\bbeta^*$ via bilevel optimization,
we use the optimized $\bbeta^*$
for one final round of function fitting
using \emph{all} of the samples.

\subsection{Oracle Analysis of INR Regularization}

To investigate the sensitivity of INR to the weight decay hyperparameters,
we performed an oracle grid search
over $\lambda_{\mathrm{Enc}}, \lambda_{\mathrm{MLP}}$
while fixing the learning rate $\tau$ and encoder scale factor $b$
at their bilevel optimized values.
The oracle optimal regularization strengths are:
\begin{equation}
\lambda_{\mathrm{Enc}}^*, \, \lambda_{\mathrm{MLP}}^*
= \argmin_{\lambda_{\mathrm{Enc}}, \, \lambda_{\mathrm{MLP}}}
\ \frac{\left\|
f_{\hat{\btheta}(\lambda_{\mathrm{Enc}}, \, \lambda_{\mathrm{MLP}};\,\tau, \, b)}
(\tilde{\z}) - f_{\mathrm{true}}(\tilde{\z}) \right\|_2}
{\left\|f_{\mathrm{true}}(\tilde{\z})\right\|_2}.
\label{eq:inr_oracle}
\end{equation}

For the validation loss based criterion, the optimal hyperparameters 
are selected by minimizing the validation loss:
\begin{equation}
\lambda_{\mathrm{Enc}}^*,\, \lambda_{\mathrm{MLP}}^*
= \argmin_{\lambda_{\mathrm{Enc}},\, \lambda_{\mathrm{MLP}}}
\ \mathcal{L}_{\cV}\paren{\hat{\btheta}(\lambda_{\mathrm{Enc}}, \lambda_{\mathrm{MLP}};\,\tau, \, b)}.
\label{eq:inr_val_select}
\end{equation}
We compared four hyperparameter selection approaches:
(1) oracle NRMSE with grid search using 100\% of samples for INR training,
(2) oracle NRMSE with grid search using 80\% of samples for training,
(3) validation loss with grid search using an 80\%/20\% train/validation split,
and (4) validation loss with Bayesian optimization (bilevel optimization)~\cite{bilevelINR_arXiv}.

\subsection{Evaluation Metric}

For both methods, we report the NRMSE on the evaluation grid:
\begin{equation}
\mathrm{NRMSE}(\hat{f}) = \frac{\| \hat{f}(\ztilt) - f_{\mathrm{true}}(\ztilt) \|_2}{\| f_{\mathrm{true}}(\ztilt) \|_2},
\label{eq:nrmse}
\end{equation}
where $\hat{f}$ denotes the reconstructed function ($f_{\hat{\bc}}$ or $f_{\hat{\btheta}}$).

\section{Experiments}

To compare the representation capacity of INR and B-spline, we fit a 2D rect function:
\begin{equation}
    f(x,y) = \mathrm{Rect}\paren{x-\frac{3}{2}}\,\mathrm{Rect}\paren{y-\frac{3}{2}},
    \label{eq:rect_analytical}
\end{equation}
using $N = 10000$ points sampled uniformly at random on $[0,3]\times[0,3]$.
Both methods are evaluated on a dense $501\times 501$ uniform grid $\ztilt$
covering $[-0.3, 3.3]\times[-0.3, 3.3]$.
Only Tikhonov regularization on weights/coefficients is applied, 
excluding additional handcrafted regularizers to isolate the representation 
capacity of each model.

For B-spline fitting, we used cubic B-splines with uniformly spaced knots
and performed an oracle grid search over the regularization 
strength $\log_{10}\lambda \in [-5, 5]$ with step size 0.1 (101 values), 
and the number of knots $M \in \{5, 10, 15, \ldots, 100\}$ (20 values).

For INR, we used the hash-encoded architecture described in~\cite{bilevelINR_arXiv}
with an 80\%/20\% train/validation split. The bilevel optimization ran 60 upper-level 
iterations with 2,000 lower-level iterations each to obtain the optimized 
hyperparameter vector $\bbeta^*$. 
For the hyperparameter sensitivity analysis, we fixed the learning rate $\tau$ 
and encoder scale factor $b$ at their bilevel optimized values and performed 
grid searches over 
$\log_{10}\lambda_{\mathrm{Enc}} \in [-4, -1]$ and $\log_{10}\lambda_{\mathrm{MLP}} \in [-9, -6]$ 
matching the bilevel optimization range, each with step size 0.1 (31 values per axis). 
The grid search was conducted using three selection criteria: oracle NRMSE with 100\% 
training samples, oracle NRMSE with 80\% training samples, and validation loss 
with 80\%/20\% train/validation split.

\section{Results}

\subsection{B-Spline Results}
\begin{figure}[!h]
    \centering
    \includegraphics[width=\linewidth]{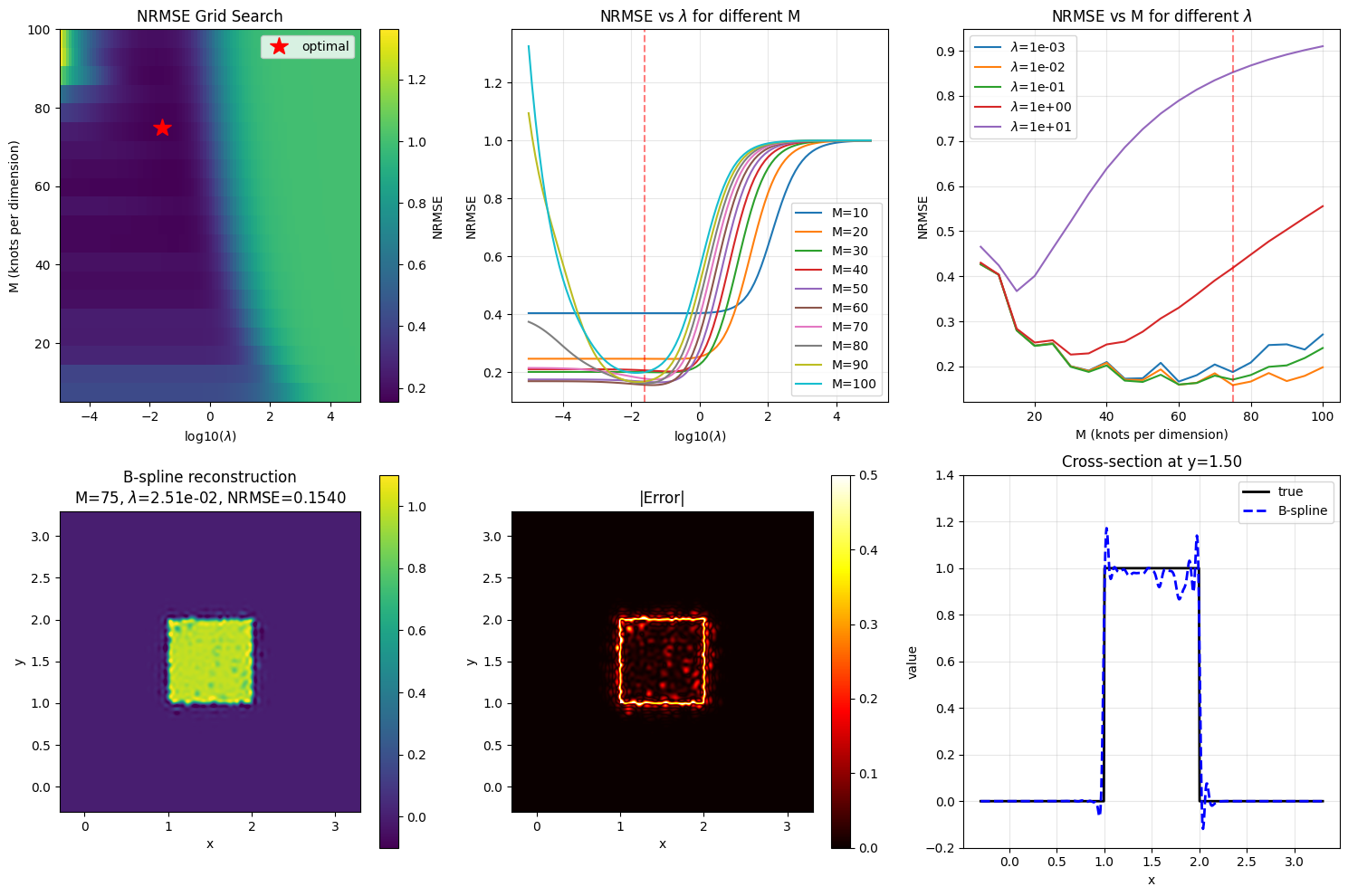}
    \caption{Cubic B-spline fitting with oracle grid search over 
    regularization strength $\lambda$ and number of knots $M$. 
    Top row: NRMSE heatmap showing the optimal at $M=75$, $\lambda=2.51\times10^{-2}$ (red star); 
    NRMSE versus $\lambda$ for different $M$; NRMSE versus $M$ for different $\lambda$. 
    Bottom row: optimal reconstruction, absolute error map, and cross-section at $y=1.50$. 
    The optimal NRMSE is 0.1540.}
    \label{fig:bspline_fit}
\end{figure}

\begin{figure}[!h]
    \centering
    \includegraphics[width=0.8\linewidth]{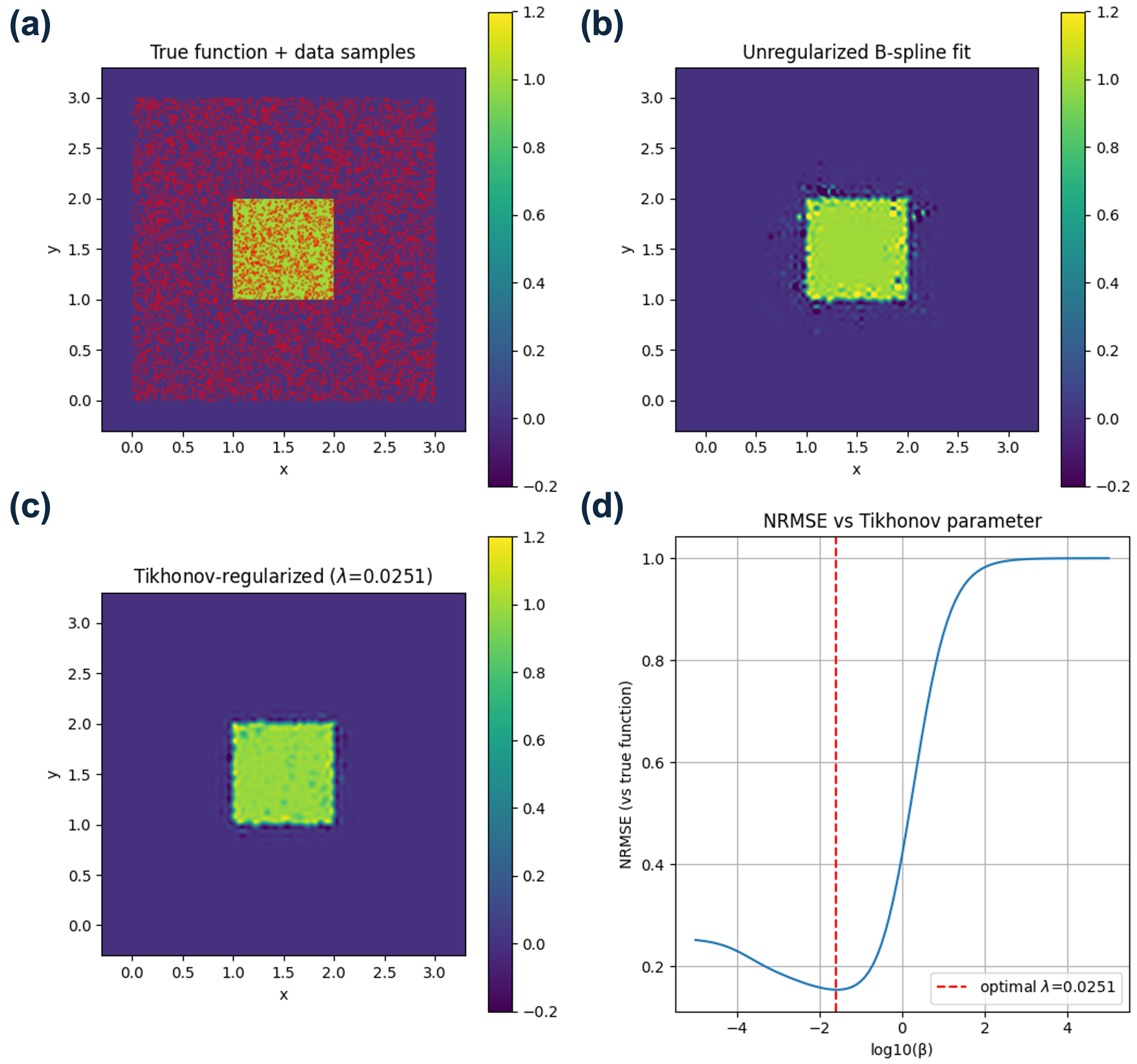}
    \caption{Effect of Tikhonov regularization on cubic B-spline fitting with $M=75$ knots. 
    (a) True function with random samples. 
    (b) Unregularized fit showing oscillatory artifacts. 
    (c) Optimally regularized fit ($\lambda=0.0251$) with reduced artifacts but increased blur. 
    (d) NRMSE versus $\lambda$, illustrating the bias-variance tradeoff.}
    \label{fig:bspline_unregularized}
\end{figure}

Fig.~\ref{fig:bspline_fit} shows the B-spline fitting results 
with oracle grid search over Tikhonov regularization strength $\lambda$ 
and number of knots $M$. The optimal hyperparameter set achieved NRMSE $= 0.154$
at $M^*=75$ and $\lambda^* = 2.51\times10^{-2}$.

The NRMSE heatmap (top left) and NRMSE vs $\lambda$ plot (top middle) reveal three 
different regimes depending on $M$. 
For $M < 50$, the B-spline model underfits the data and Tikhonov regularization 
provides no improvement. 
For $M \in [50, 80)$, regularization near $\lambda = 10^{-2}$ reduces NRMSE compared 
to the unregularized case (e.g., by 0.09 for $M=75$). 
For $M \geq 80$, the model overfits. While regularization helps, 
the minimum NRMSE remains higher than the optimal at $M=75$. 
The NRMSE versus $M$ curves (top right) further shows 
that $\lambda \approx 10^{-2}$ consistently produces the 
lowest error across different knot counts.

Even for the setting that achieved the lowest NRMSE,
the optimized B-spline reconstruction (bottom row)
exhibits noticeable artifacts near the 
discontinuities and residual oscillations in the flat-top region,
as shown in the cross-section at $y=1.50$.

Fig.~\ref{fig:bspline_unregularized} compares the unregularized 
and optimally regularized B-spline fits at $M=75$. 
The unregularized fit (b) shows pronounced oscillatory artifacts 
near the rectangle edges due to overfitting. 
Tikhonov regularization (c) suppresses these artifacts but 
introduces visible blurring. 
The NRMSE curve (d) shows a general trend of Tikhonov regularization 
in B-spline fitting: insufficient regularization ($\lambda < 10^{-3}$) 
results in overfitting, whereas excessive regularization ($\lambda > 10^{0}$) 
causes over-smoothing.

\subsection{INR Results}

\begin{figure}[!h]
    \centering
    \includegraphics[width=\linewidth]{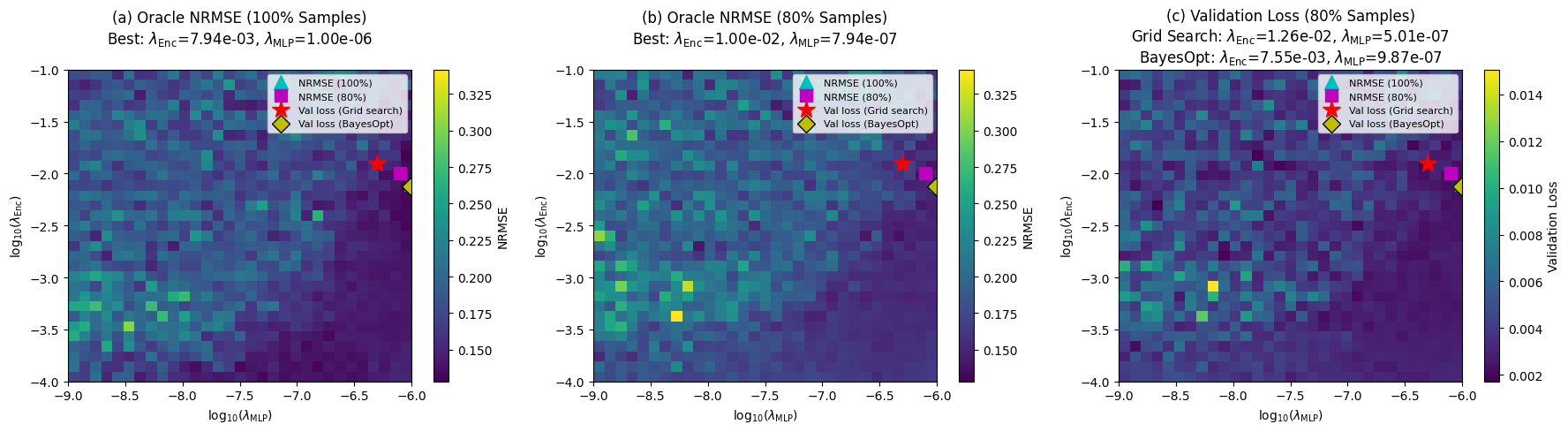}
    \caption{INR hyperparameter sensitivity analysis over encoder 
    and MLP weight decay ($\lambda_{\mathrm{Enc}}$, $\lambda_{\mathrm{MLP}}$). 
    (a) Oracle NRMSE with 100\% training samples. 
    (b) Oracle NRMSE with 80\% training samples. 
    (c) Validation loss with 80\%/20\% train/validation split. 
    Markers indicate optima: cyan triangle (100\% oracle), 
    magenta square (80\% oracle), 
    red star (validation grid search), 
    yellow diamond (BayesOpt).}
    \label{fig:inr_heatmaps}
\end{figure}

Fig.~\ref{fig:inr_heatmaps} shows the INR hyperparameter sensitivity analysis 
over the encoder and MLP weight decay parameters ($\lambda_{\mathrm{Enc}}$, 
$\lambda_{\mathrm{MLP}}$). The oracle NRMSE heatmap with 100\% training samples 
(a) yields the optimal weight decay pair 
$(\lambda_{\mathrm{Enc}}, \lambda_{\mathrm{MLP}}) = (7.94\times10^{-3}, 1.00\times10^{-6})$. 
Using 80\% training samples (b) produces a similar optimum 
at $(1.00\times10^{-2}, 7.94\times10^{-7})$, 
indicating that the reduced training set has only a modest effect on 
the optimal hyperparameters in this task.

When using validation loss as the selection criterion with grid search (c), 
the optimum shifts to $(1.26\times10^{-2},\ 5.01\times10^{-7})$, 
which is further from the oracle solution. 
However, the bilevel optimization scheme using Bayesian optimization (BayesOpt) 
finds a superior solution at $(7.55\times10^{-3},\ 9.87\times10^{-7})$, 
closely approximating the oracle optimum despite having no access to the ground truth 
and limited iterations (60 upper-level iterations
compared to 961 grid search evaluations). 

\begin{figure*}[ht]
    \centering
    \includegraphics[width=0.8\linewidth]{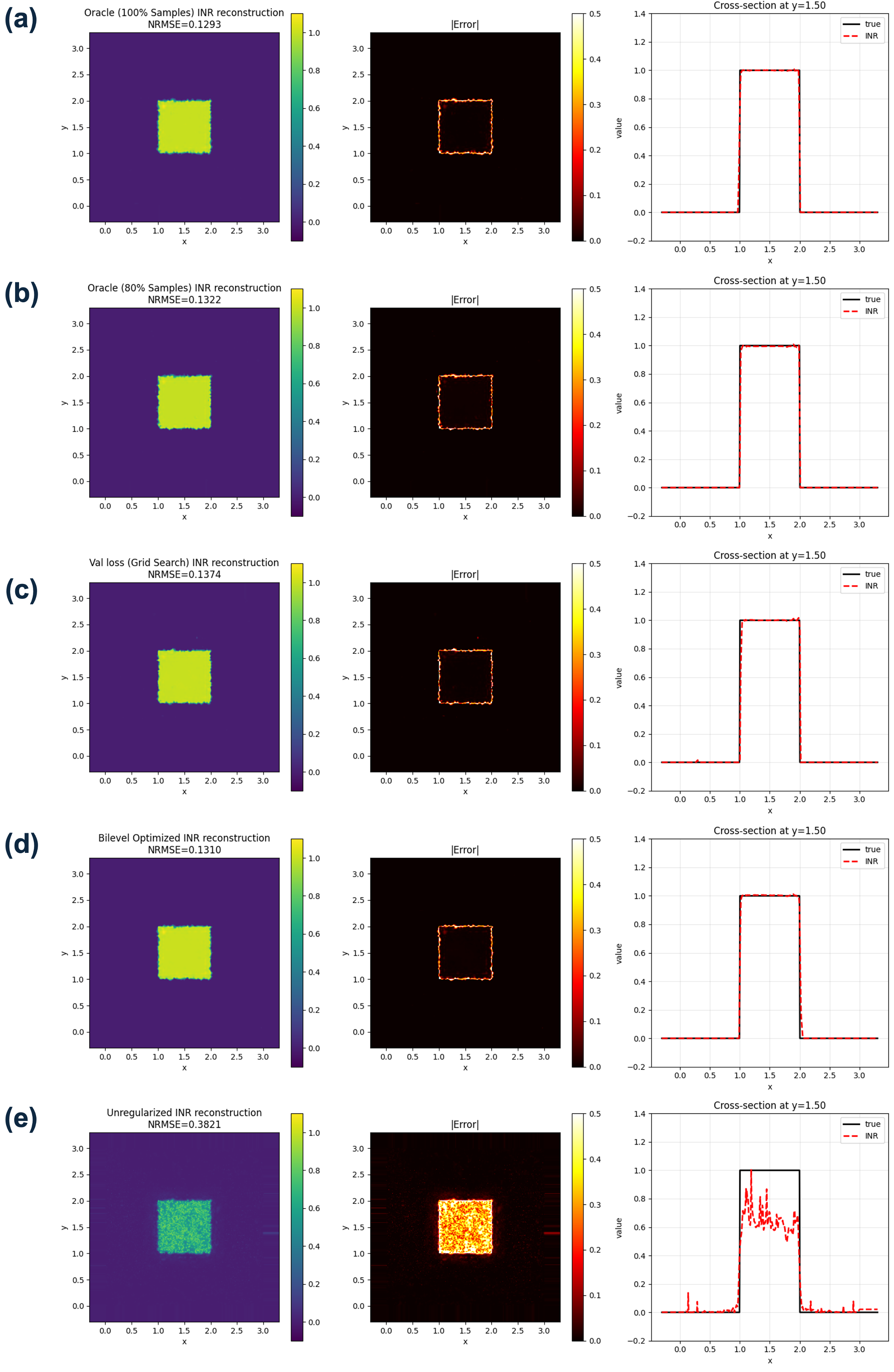}
    \caption{INR reconstruction results under different hyperparameter selection criteria. 
    Each row shows the reconstruction, absolute error map, and cross-section at $y=1.50$. 
    (a) Oracle grid search with 100\% samples (NRMSE $= 0.129$). 
    (b) Oracle grid search with 80\% samples (NRMSE $= 0.132$). 
    (c) Validation loss grid search with 80\%/20\% split (NRMSE $= 0.137$). 
    (d) Bilevel optimization with 80\%/20\% split (NRMSE $= 0.131$).
    (e) Unregularized fitting (NRMSE $= 0.382$).}
    \label{fig:inr_fit}
\end{figure*}

Fig.~\ref{fig:inr_fit} compares INR reconstructions under different 
hyperparameter selection strategies. 
Among the four regularized approaches (a--d), oracle grid search with 100\% samples 
achieved the lowest NRMSE of 0.129,
representing the best achievable performance. 
The bilevel optimized INR (d) achieves NRMSE $= 0.131$, closely matching the oracle 
and outperforming both the 80\% oracle grid search (NRMSE $= 0.132$) and the 
validation loss grid search (NRMSE $= 0.137$). 
Notably, all four regularized INR results outperformed
the optimal B-spline result (NRMSE $= 0.154$).

Visually, all regularized INR reconstructions exhibit minimal artifacts 
in the flat-top region of the rectangle, 
with residual errors mainly appearing near the discontinuities. 
The cross-section plots at $y=1.50$ confirm sharper edge transitions and reduced 
ringing compared to the B-spline results in Fig.~\ref{fig:bspline_fit}.

The unregularized INR result in Fig.~\ref{fig:inr_fit}
demonstrates the critical importance of weight regularization for INR. 
Despite achieving near-zero training loss,
the overparameterized nonlinear model severely overfits without regularization,
resulting in NRMSE $= 0.3821$ with pronounced artifacts:
underestimation within the rectangle region and spurious oscillations elsewhere. 
This confirms that, without proper regularization,
the expressive capacity of hash-encoded INR leads to poor generalization
from sparse samples.

\section{Discussion and Conclusions}

This work presents preliminary experiments
comparing INR and finite-series B-spline representations 
for continuous function fitting from sparse random samples. 
Both methods used only Tikhonov regularization on their respective weights/coefficients,
isolating the comparison to representation capacity without additional handcrafted priors.

\subsection{Comparison of Representation Capacity}

The results demonstrate that, under oracle hyperparameter selection, 
the hash-encoded INR (NRMSE $= 0.129$) outperformed
the cubic B-spline (NRMSE $= 0.154$)
for fitting a 2D rect function.
The INR reconstructions exhibited sharper edge transitions and fewer oscillatory artifacts 
in the flat-top region compared to B-spline, as shown in the cross-section plots.

However, the improved performance comes with tradeoffs. 
The INR contains substantially more parameters than the B-spline model, 
making it less parameter-efficient. 
Additionally, the B-spline formulation is more interpretable, and
its NRMSE heatmap is also smoother than the INR heatmaps.
The ``rougher'' INR landscape may be due to the random small weight initialization 
used for each fit,
which introduces the only source of stochasticity
in this toy experiment.
By contrast,
in the MRI reconstruction experiments~\cite{bilevelINR_arXiv},
repeated runs with different initializations produced nearly identical reconstructions, 
suggesting that weight initialization has a smaller practical impact in that setting.
A more thorough theoretical analysis of INR representation and its implicit biases 
remains an active research direction.

\subsection{Effect of Tikhonov Regularization}

Both methods benefit from Tikhonov regularization, but the effect differs due 
to their model differences. 
For B-spline fitting, regularization directly penalizes coefficient magnitudes, 
introducing a bias-variance tradeoff where reduced oscillations come at the 
cost of increased blurring (Fig.~\ref{fig:bspline_unregularized}).

For INR, the encoder-decoder architecture enables a different regularization scheme. 
Weight penalty on the encoder ($\lambda_{\mathrm{Enc}}$) directly regularizes 
the multiresolution feature representation, 
while the smaller MLP regularization ($\lambda_{\mathrm{MLP}}$) only mildly constrains 
the decoded continuous function. 
The final linear layer can rescale outputs, 
partially compensating for magnitude shrinkage from regularization of the encoder.
This, combined with the much richer nonlinear function class provided by the 
overparameterized model, may explain why INR achieved lower NRMSE
with Tikhonov regularization alone, 
without requiring additional image-domain priors.

The unregularized INR result (NRMSE $= 0.382$)
further emphasizes the importance of regularization  
for the overparameterized hash-encoded architecture. 
Without weight regularization,
the model overfit severely despite achieving near-zero training loss.

\subsection{Efficacy of Bilevel Optimization}

The hyperparameter sensitivity analysis
demonstrates that bilevel optimization using Bayesian optimization
can effectively approximate oracle performance
like the accelerated MRI reconstruction task. 
The bilevel optimized INR (NRMSE $= 0.131$) closely matched
the 100\% oracle result (NRMSE $= 0.129$) and outperformed
validation-based grid search (NRMSE $= 0.137$),
despite operating in continuous hyperparameter space with only 
60 upper-level iterations compared to 961 grid search evaluations.

This result supports the use of bilevel optimization for automatic 
hyperparameter selection when ground truth is 
unavailable, as is the case in practical MRI reconstruction.

\subsection{Conclusions}

In summary, this preliminary study provides empirical evidence that hash-encoded INR 
with Tikhonov regularization can outperform cubic B-spline for fitting discontinuous 
functions from sparse samples, and that bilevel optimization
can effectively select
near-optimal hyperparameters without ground truth access.
Further study on extending the experiments to more complex tasks, 
such as realistic MR reconstruction settings,
remains a future direction.
It would also be interesting to investigate
fitting a smooth function that can be represented perfectly by cubic B-splines.

\section*{Acknowledgment}
This study was supported in part by NIH grants R37CA263583, R01CA284172, Siemens Healthineers and Cook Medical.

\bibliographystyle{IEEEtran}
\bibliography{mybib}
\end{document}

%% file: macros.tex
\long\def\comment#1{}

\newcommand{\xmath}[1] {\ensuremath{#1}\xspace}
\newcommand{\blmath}[1] {\xmath{\bm{#1}}}

\newcommand{\y}{\blmath{y}}
\newcommand{\z}{\blmath{z}}

\newcommand{\reals} {\xmath{\mathbb{R}}}

\newcommand{\bc}{\boldsymbol{c}}

\newcommand{\bI}{\boldsymbol{I}}

\newcommand{\bbeta}{\boldsymbol{\beta}}

\newcommand{\btheta}{\xmath{\bm{\theta}}}

\newcommand{\bPhi}{\boldsymbol{\Phi}}

\newcommand{\paren}[1]{\left( #1 \right)}

\long\def\red#1{\bgroup\color{red}#1\egroup}
\long\def\purple#1{\bgroup\color{purple}#1\egroup}

\newcommand{\cT}{\xmath{\mathcal{T}}}
\newcommand{\cV}{\xmath{\mathcal{V}}}

\long\def\blue#1{\bgroup\color{mich-blue-high}#1\egroup}
\long\def\red#1{\bgroup\color{red}#1\egroup}

\makeatletter
\newcommand{\pictslash}[2]{%
  \vcenter{%
    \sbox0{$\m@th#1\varobslash$}\dimen0=.55\wd0
    \hbox to\wd 0{%
      \hfil\pictslash@aux#2\hfil
    }%
  }%
}
\newcommand{\pictslash@aux}[2]{%
    \begin{picture}(\dimen0,\dimen0)
    \roundcap
    \put(0,#1\dimen0){\line(1,#2){\dimen0}}
    \end{picture}%
}
\makeatother

\makeatletter
\newcommand*{\rom}[1]{\expandafter\@slowromancap\romannumeral #1@}
\makeatother

\newcommand{\RNum}[1]{\uppercase\expandafter{\romannumeral #1\relax}}
\newcommand{\vecb}[1]{\vec{\bm{#1}}}
\newcommand{\ztilt}{\tilde{\bm{z}}}

\usepackage{bm,xspace}
\usepackage{amsmath, esvect}

\DeclareMathOperator*{\argmin}{arg\,min}

\long\def\red#1{\bgroup\color{red}#1\egroup}

\newif\ifshowrevisions
\showrevisionstrue

%% file: mybib.bib
@inproceedings{nerf,
 title={{NeRF}: Representing Scenes as Neural Radiance Fields for View Synthesis},
 author={Ben Mildenhall and Pratul P. Srinivasan and Matthew Tancik and Jonathan T. Barron and Ravi Ramamoorthi and Ren Ng},
 year={2020},
 booktitle={Proc. ECCV},
}

@inproceedings{FourierFeautreNet,
    title={{Fourier} Features Let Networks Learn High Frequency Functions in Low Dimensional Domains},
    author={Matthew Tancik and Pratul P. Srinivasan and Ben Mildenhall and Sara Fridovich-Keil and Nithin Raghavan and Utkarsh Singhal and Ravi Ramamoorthi and Jonathan T. Barron and Ren Ng},
    booktitle= {Proc. NIPS},
    year={2020}
}

@article{instantngp,
    author = {Thomas M\"uller and Alex Evans and Christoph Schied and Alexander Keller},
    title = {Instant Neural Graphics Primitives with a Multiresolution Hash Encoding},
    journal = {ACM Trans. Graph.},
    issue_date = {July 2022},
    volume = {41},
    number = {4},
    month = jul,
    year = {2022},
    pages = {102:1--102:15},
    articleno = {102},
    numpages = {15},
    doi = {10.1145/3528223.3530127},
    publisher = {ACM},
    address = {New York, NY, USA}
}

@inproceedings{siren,
                author = {Sitzmann, Vincent
                          and Martel, Julien N.P.
                          and Bergman, Alexander W.
                          and Lindell, David B.
                          and Wetzstein, Gordon},
                title = {Implicit Neural Representations
                          with Periodic Activation Functions},
                booktitle = {Proc. NIPS},
                year={2020}
            }

@ARTICLE{IMJENSE,
  author={Ruimin Feng and Qing Wu and Jie Feng and Huajun She and Chunlei Liu and Yuyao Zhang and Hongjiang Wei},
  journal={IEEE Trans. on Med. Imag.}, 
  title={{IMJENSE}: Scan-Specific Implicit Representation for Joint Coil Sensitivity and Image Estimation in Parallel {MRI}}, 
  year={2024},
  volume={43},
  number={4},
  pages={1539-1553},
  doi={10.1109/TMI.2023.3342156}}

@ARTICLE{NeRP,
  author={Shen, Liyue and Pauly, John and Xing, Lei},
  journal={IEEE Trans. Neural Netw. and Learn. Syst.}, 
  title={{NeRP}: Implicit Neural Representation Learning With Prior Embedding for Sparsely Sampled Image Reconstruction}, 
  year={2024},
  volume={35},
  number={1},
  pages={770-782},
  doi={10.1109/TNNLS.2022.3177134}}

@misc{adam,
      title={Adam: A Method for Stochastic Optimization}, 
      author={Diederik P. Kingma and Jimmy Ba},
      year={2017},
      eprint={1412.6980},
      archivePrefix={arXiv},
      primaryClass={cs.LG},
      url={https://arxiv.org/abs/1412.6980}, 
}

@article{spatiotemporal_inr,
  author={Feng, Jie and Feng, Ruimin and Wu, Qing and Shen, Xin and Chen, Lixuan and Li, Xin and Feng, Li and Chen, Jingjia and Zhang, Zhiyong and Liu, Chunlei and Zhang, Yuyao and Wei, Hongjiang},
  journal={IEEE Trans. Med. Imag.}, 
  title={Spatiotemporal implicit neural representation for unsupervised dynamic {MRI} reconstruction}, 
  year={2025},
  volume={},
  number={},
  pages={1-1}}

@article{MBIR_MRI,
  author  = {Fessler, Jeffrey A.},
  title   = {Model-Based Image Reconstruction for {MRI}},
  journal = {IEEE Signal Process. Mag.},
  year    = {2010},
  volume  = {27},
  number  = {4},
  pages   = {81--89},
  doi     = {10.1109/MSP.2010.936726},
}

@book{wahba1990spline,
  title={{Spline Models for Observational Data}},
  author={Wahba, Grace},
  year={1990},
  publisher={SIAM}
}

@ARTICLE{splineSPIM,
  author={Unser, Michael},
  journal={IEEE Signal Process. Mag.}, 
  title={Splines: a perfect fit for signal and image processing}, 
  year={1999},
  volume={16},
  number={6},
  pages={22-38},
  doi={10.1109/79.799930}}

@book{mallat1999wavelet,
  title={A wavelet tour of signal processing},
  author={Mallat, St{\'e}phane},
  year={1999},
  publisher={Academic Press}
}

@book{wendland2005scattered,
  title={{Scattered Data Approximation}},
  author={Wendland, Holger},
  year={2005},
  publisher={Cambridge University Press},
  address={Cambridge},
  volume={17},
  doi={10.1017/CBO9780511617539}
}

@Article{kerbl3Dgaussians,
      author       = {Kerbl, Bernhard and Kopanas, Georgios and Leimk{\"u}hler, Thomas and Drettakis, George},
      title        = {{3D Gaussian Splatting for Real-Time Radiance Field Rendering}},
      journal      = {ACM Trans. Graphics},
      number       = {4},
      volume       = {42},
      month        = {July},
      year         = {2023},
      url          = {https://repo-sam.inria.fr/fungraph/3d-gaussian-splatting/}
}

@inproceedings{park2019deepsdf,
  title={{DeepSDF}: Learning Continuous Signed Distance Functions for Shape Representation},
  author={Park, Jeong Joon and Florence, Peter and Straub, Julian and Newcombe, Richard and Lovegrove, Steven},
  booktitle={Proc. CVPR},
  pages={165--174},
  year={2019}
}

@misc{bilevelINR_arXiv,
      title={Bilevel Optimized Implicit Neural Representation for Scan-Specific Accelerated MRI Reconstruction}, 
      author={Hongze Yu and Jeffrey A. Fessler and Yun Jiang},
      year={2025},
      eprint={2502.21292},
      archivePrefix={arXiv},
      primaryClass={eess.IV},
      url={https://arxiv.org/abs/2502.21292}, 
}

@article{lyu2026rapid,
  title={Rapid whole brain motion-robust mesoscale in-vivo {MR} imaging using multi-scale implicit neural representation},
  author={Lyu, Jun and Ning, Lipeng and Consagra, William and Liu, Qiang and Rushmore, Richard J. and Bilgic, Berkin and Rathi, Yogesh},
  journal={Med. Imag. Analysis},
  volume={107},
  pages={103830},
  year={2026},
  publisher={Elsevier},
  doi={10.1016/j.media.2025.103830}
}

@article{barron2021mipnerf,
    title={{Mip-NeRF}: A Multiscale Representation 
           for Anti-Aliasing Neural Radiance Fields},
    author={Jonathan T. Barron and Ben Mildenhall and 
            Matthew Tancik and Peter Hedman and 
            Ricardo Martin-Brualla and Pratul P. Srinivasan},
    journal={ICCV},
    year={2021}
}

@inproceedings{fridovich2022plenoxels,
  title={Plenoxels: Radiance Fields without Neural Networks},
  author={Fridovich-Keil, Sara and Yu, Alex and Tancik, Matthew and Chen, Qinhong and Recht, Benjamin and Kanazawa, Angjoo},
  booktitle={Proc. CVPR},
  pages={5501--5510},
  year={2022}
}
